\documentclass[namedreferences,journalpaper,10pt,final,twoside,onecolumn]{SolarPhysicsMod}
\usepackage[optionalrh,solaenum]{spr-sola-addons} 
\usepackage{graphicx}                    
\usepackage{color}                       
\usepackage{url}                         
\usepackage{cases}
\usepackage{marvosym}

\usepackage{mathptmx}

\usepackage{solaheader}	
\newcommand{\solareceiveddate}{7 January 2010} 
\newcommand{\solaaccepteddate}{21 March 2010}

\newcommand{\degree}{\ensuremath{^\circ}}

\begin{document}

\begin{article}

\begin{opening}

\title{Signatures of Emerging Subsurface Structures in Acoustic Power Maps}

\author{T.~Hartlep \sep A.G.~Kosovichev \sep J.~Zhao \sep N.N.~Mansour}

%
\runningauthor{Hartlep {\sl et~al.}}
\runningtitle{Signatures of Emerging Subsurface Structures in Acoustic Power Maps}

 \institute{
  T.~Hartlep (\,{\large\Letter}) \sep A.G.~Kosovichev \sep J.~Zhao \\
  W.W.~Hansen Experimental Physics Laboratory, Stanford University, Stanford, CA, USA \\
  email: \url{thartlep@sun.stanford.edu} \\
  \vspace*{1em}
  N.N.~Mansour \\
  NASA Ames Research Center, Moffett Field, CA, USA
  }

\begin{abstract}

We show that under certain conditions, subsurface structures in the solar interior can alter the average acoustic power observed at the photosphere above them.
By using numerical simulations of wave propagation, we show that this effect is large enough for it to be potentially used for detecting emerging active regions before they appear on the surface.
In our simulations, simplified subsurface structures are modeled as regions with enhanced or reduced acoustic wave speed.
We investigate the dependence of the acoustic power above a subsurface region on the sign, depth, and strength of the wave speed perturbation.
Observations from the {\sl Solar and Heliospheric Observatory/Michelson Doppler 
Imager} (SOHO/MDI) prior and during the emergence of NOAA active region 10488 are used to test the use of acoustic power as a potential precursor of magnetic flux emergence.

\end{abstract}

%

\end{opening}


\section{Motivation and Objectives}

The complex dynamics that lead to the emergence of active regions on the Sun are poorly understood. 
One possibility is that magnetic structures (flux tubes, {\sl etc.\/}) rising from below the surface by self induction and convection cause the formation of active regions and sunspots on the solar surface.
For space-weather forecasting, one would like to detect large subsurface structures before they reach the surface.
The goal of this study is to investigate whether perturbations in the wave propagation speed associated with subsurface structures could affect the acoustic power observed at the solar surface above them.
Possible mechanisms for this are wave reflection, scattering, or diffraction. 
If the power variations are above the solar noise level, this effect may be used to detect emerging subsurface structures before an active region appears on the solar surface.

\section{Numerical method}

\subsection{Simulation code}

In the following, we give a brief overview of the 3D numerical simulation code used in this study.
For more details, the reader is referred to \inlinecite{har05} and \inlinecite{2008ApJ...689.1373H}.

We model solar acoustic oscillations in a spherical domain using linearized Euler equations and consider a static background in which only localized variations of the sound speed are taken into account.
In this simple model, flows and magnetic fields are not included.
The oscillations are assumed to be adiabatic, and are driven by randomly forcing density perturbations near the surface.
For the unperturbed background model of the Sun, we use standard solar model S of~\inlinecite{1996Sci...272.1286C} matched to a model of the chromosphere~\cite{1981ApJS...45..635V}.
Localized sound-speed perturbations of various sizes are added as simplified subsurface structures.
Non-reflecting boundary conditions are applied at the upper boundary by means of an absorbing buffer layer with a damping coefficient that is set to zero in the interior and increases smoothly into the buffer layer.
Perturbations of the gravitational potential are neglected, and the adiabatic approximation is used.
In order to make the linearized equations convectively stable, we also neglect the entropy gradient of the background model.
The calculations show that this assumption does not significantly change the propagation properties of acoustic waves including their frequencies, except for the acoustic cut-off frequency, which is slightly reduced.
Because of this, high mode frequencies above approximately 3.5~mHz have reduced amplitudes in the simulation compared to solar observations.
For comparison, other authors have modified the solar model including its sound speed profile~({\sl e.g.\rm}, \opencite{2006ApJ...648.1268H}; \opencite{2007ApJ...666L..53P}) in order to stabilize the simulations.
In those cases, the oscillation mode frequencies may differ significantly from the frequencies in the real Sun.

For numerical discretization, a Galerkin scheme is applied where spherical harmonic functions are used for the angular dependencies, and fourth-order B-splines~\cite{Loulou97,1999JCoPh.151..757K} are used for the radial direction.
The radial resolution of the B-spline method is varied proportionally to the speed of sound, {\sl i.e.} the generating knot points are closely spaced near the surface (where the sound speed is small), and are coarsely spaced in the deep interior (where the sound speed is large).
In the simulations presented in this paper  three hundred B-splines are used in the radial direction with a know spacing that varies between approximately 100~km and 7~Mm.
The simulations employ spherical harmonics of angular degree $\ell$ from 0 to 170.
Two-thirds dealiasing is used.
A staggered Yee scheme~\cite{1966ITAP...14..302Y} is used for time integration, with a time step of two seconds.
Simulations were run for at least 17 hours of solar time.

The simulation code has been validated and successfully used for testing helioseismic far-side imaging by simulating the effects of model sunspots on the acoustic field~\cite{2008ApJ...689.1373H,2009SoPh..258..181I}, and for validating time\,--\,distance helioseismic measurements of tachocline perturbations~\cite{2009ApJ...702.1150Z}.

\subsection{Subsurface structures}

For this study, we consider simple subsurface structures in which the sound speed $(c)$ differs from the sound speed of the standard solar model $(c_o)$ in the following fashion:
\begin{equation}
	\left( \frac{c}{c_o} \right)^2 = 1 + A
		\left\{ \begin{array}{ll}
			\frac{1}{4} \left( 1+\cos \pi \frac{z-z_s}{d_s}  \right)  \left( 1 + \cos \pi \frac{\alpha}{\alpha_d} \right) & \mbox{for~} |\frac{\alpha}{\alpha_d}| \le 1, |\frac{z-z_s}{d_s}| \le 1; \\
			0 & \mbox{otherwise},
      		\end{array} \right.
\end{equation}  
where $\alpha$ is the angular distance from the center of a subsurface structure, and $z$ is the depth from the photosphere.
The photosphere in the simulation is at a fixed radius of approximately 695.99~Mm (the solar radius in standard solar model~S).
The quantities $z_s$, $d_s$, $\alpha_d$, and $A$ denote the central depth at which the sound-speed perturbation is largest, the size in radial (depth) direction, the angular (horizontal) size, and the amplitude of the sound-speed variation, respectively.

\begin{figure}
	\centering
	\vspace{0.25cm}
	\includegraphics[width=5cm]{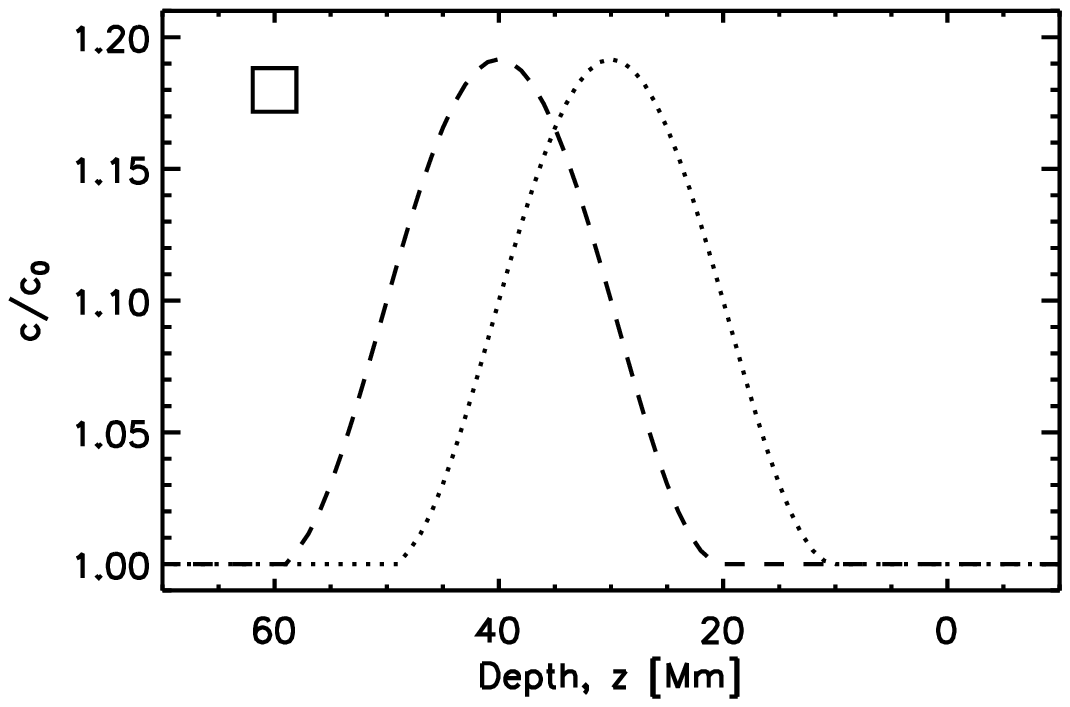}
	\includegraphics[width=5cm]{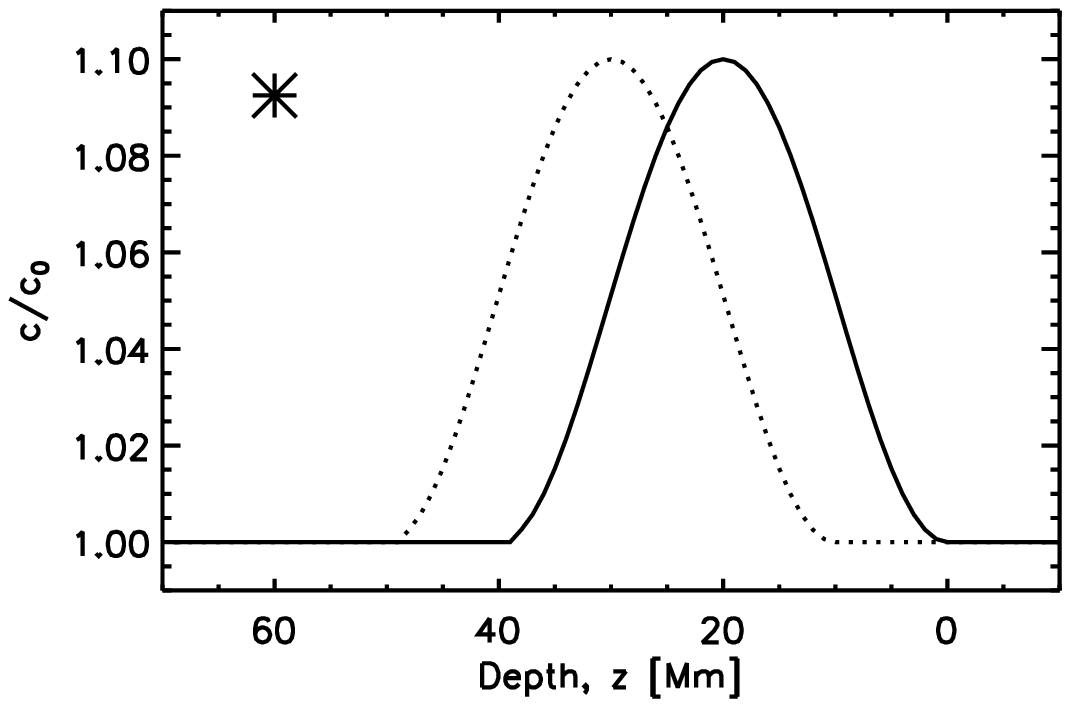}
	\\
	\includegraphics[width=5cm]{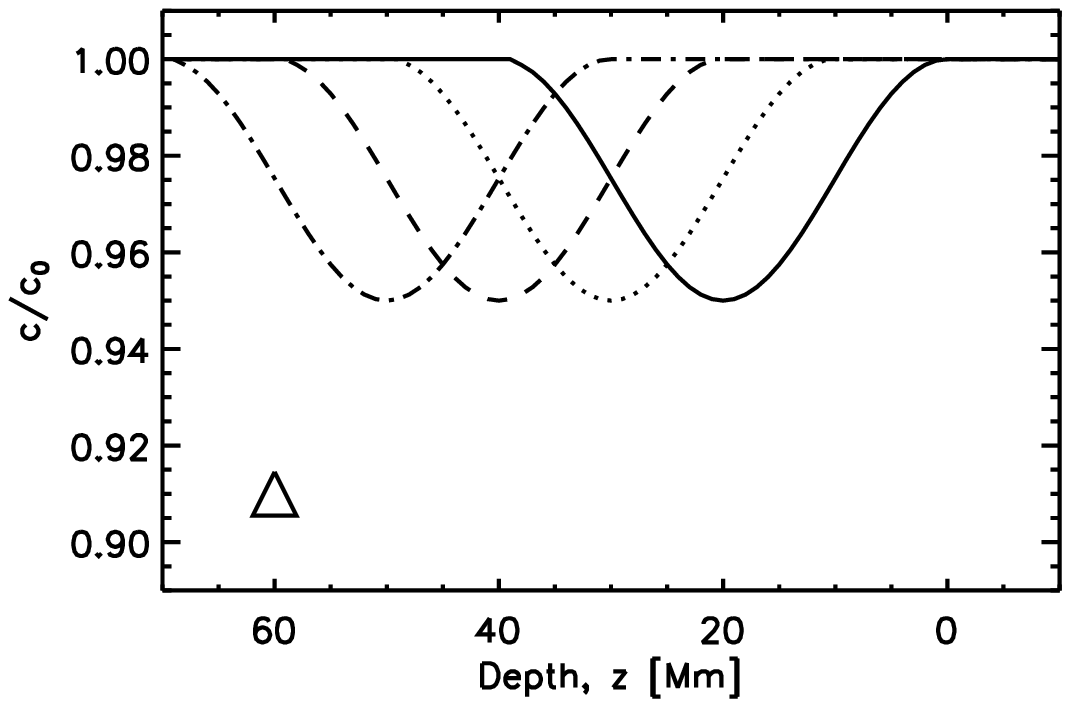}
	\includegraphics[width=5cm]{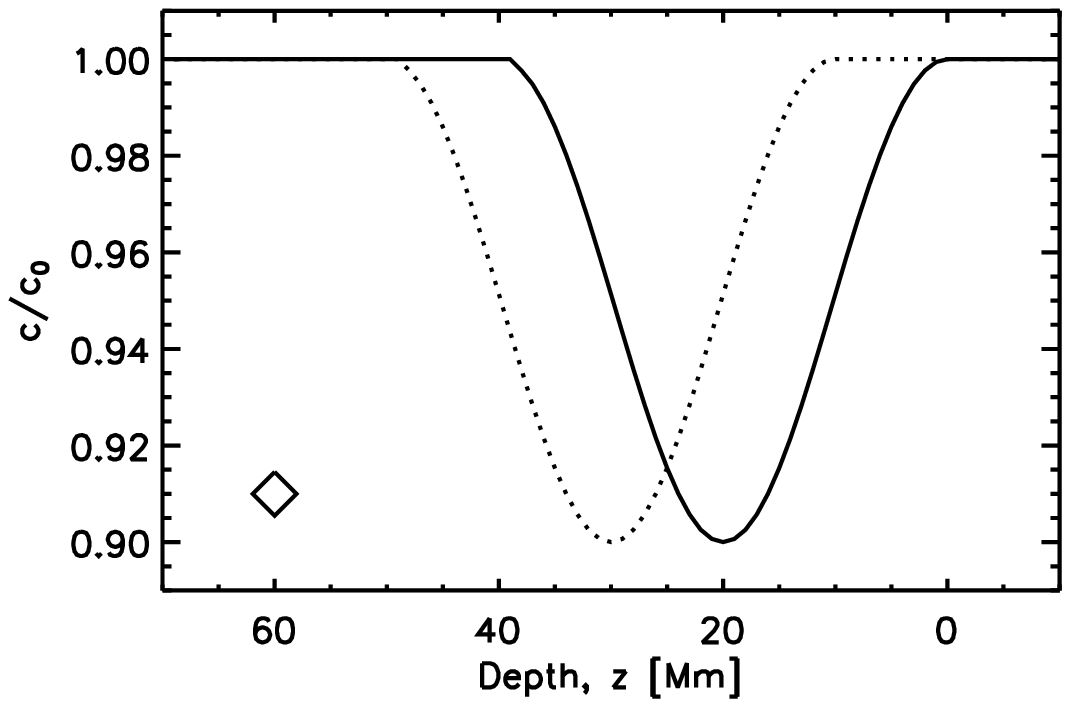}
	\vspace{0.25cm}
	\caption{Radial profiles of the sound-speed perturbation of all model subsurface regions used in this study. Depth $(z)$ is measured downwards from the solar photosphere.
	Solid, dotted, dashed, and dash-dotted lines denote perturbations with a central depth of 20, 30, 40, and 50~Mm, respectively.
	Each panel also contains a symbol which is used in Figure~\ref{Fig:Simulation:PowerVsDepth} to distinguish the four perturbations amplitudes.
	}
	\label{Fig:Simulation:Profiles}
\end{figure}
%

\section{Results}

\subsection{Simulations}

Simulations have been performed for subsurface regions of three angular sizes ($\alpha_d=3.7\degree, 7.4\degree$, and $14.8\degree$) corresponding to horizontal radii of the perturbation of 45, 90, and 180~Mm, respectively.
Four perturbation depths (20, 30, 40, and 50~Mm), and four amplitudes were considered.
The extent in radial direction ($d_s$) was 20~Mm for all cases. 
Radial profiles of all cases considered in this study are shown in Figure~\ref{Fig:Simulation:Profiles}.
In order to reduce the number of separate calculations, each simulation contained three subsurface regions which were placed as far apart as possible in order to avoid interference.

\begin{figure}
	\centering
	\includegraphics[width=8cm]{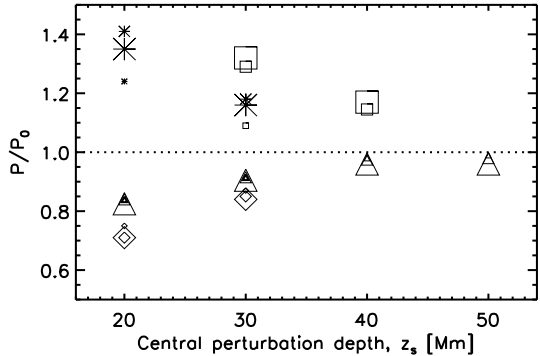}
	\caption{Ratio between the average acoustic power in the center above a simulated subsurface region and away from the region as a function of the central depth of the perturbation. The symbols are the same as in Figure~\ref{Fig:Simulation:Profiles} denoting subsurface regions with sound-speed variations of different signs and strengths: 19\% peak sound-speed increase ({\sl squares\/}), 10\% increase ({\sl stars\/}), 5\% decrease ({\sl triangles\/}), and 10\% decrease ({\sl diamonds\/}). The horizontal size of the regions (45, 90, and 180~Mm radius) are indicated by the different size of the symbols used from small to large, respectively.}
	\label{Fig:Simulation:PowerVsDepth}
\end{figure}                 

\begin{figure}
  \centering
  \includegraphics[width=10cm]{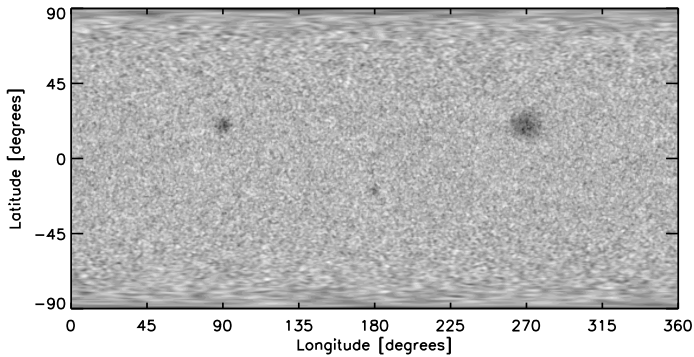}
  \vspace*{3mm}
  \caption{Acoustic power map at 300~km above the photosphere from a simulation with three differently-sized subsurface regions each with a maximum of 10\% reduction in sound speed at a depth of 20~Mm. The quantity shown is the square of the radial velocity, averaged over 824~min, with low and high values indicated in dark and bright, respectively. A model subsurface region of 45~Mm radius is located at a latitude of $-20\degree$ and longitude of $180\degree$. Two larger regions of 90 and 180~Mm radius are located at a latitude of $+20\degree$, and longitudes of $90\degree$ and $270\degree$, respectively.
The map has a resolution of approximately 0.703\degree~per pixel both in latitude and longitude.
  }
  \label{Fig:Powermap}
\end{figure}

A compilation of the main simulation results is given in Figure~\ref{Fig:Simulation:PowerVsDepth}.
The diagram shows the ratio between the acoustic power in the center above a subsurface region and far away from such regions (in a ``quiet Sun'' zone).
The acoustic power is measured by averaging the square of the radial velocity.
The horizontal velocity components of the waves considered here (low to medium spherical harmonic degrees) can be neglected at the solar surface.
It is evident that regions with positive sound-speed perturbation cause an increase in the acoustic power, and that negative perturbations cause a decrease. 
Also, stronger sound-speed perturbations produce stronger power changes.
Except for one outlier, the maximum power change depends only weakly on the horizontal size of the subsurface region and is generally larger for the larger regions.
As one would expect, the power change is weaker for deeper perturbations.
It should be noted that different frequency components show different amounts of acoustic power increase or decrease even though the qualitative behavior is the same.
For the case from Figure~\ref{Fig:Powermap}, for example, a maximum reduction in power of about 35\% is found for frequencies between 2 and 4 mHz, but the reduction is less at lower frequencies (25\%) and even less at frequencies above 4 mHz (15\%).

The simulations were carried out for static subsurface perturbations that stay at a fixed depth for several hours. 
In reality, emerging flux regions may emerge relatively fast, and can be considered approximately static only if we work with rather short observation time series.
In many of the cases considered in the simulations, the power change is relatively large, often close to or greater than 10\%.
Thus, if such structures exist, they should be measurable even from such short time series.

%

\subsection{Observations}

\begin{figure}
	\centering
	\includegraphics[width=11.0cm]{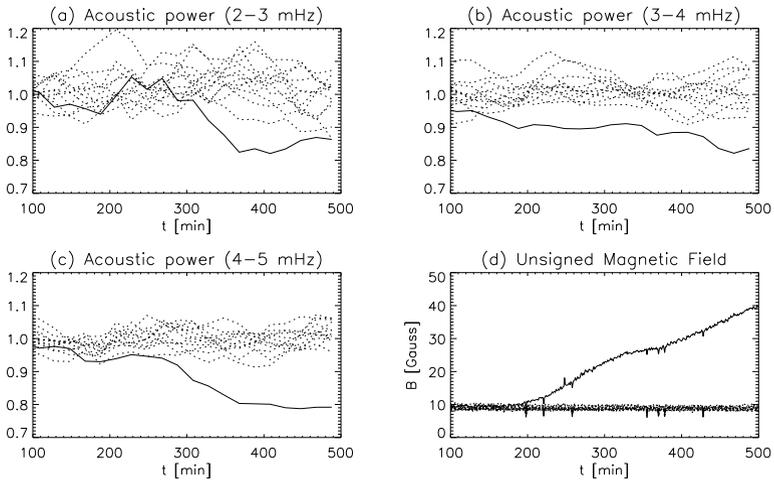}
	\caption{Time series of the observed acoustic power during the emergence of NOAA active region 10488 on 26 October 2003. Shown is, for three different frequency intervals, the power density of Doppler velocities (one-minute cadence) measured in a patch where the active region is emerging ({\sl solid line\/}) and in quiet-Sun patches ({\sl dotted lines\/}), averaged over the preceding 128~minutes. The values are normalized by the power density at the same disk location but 24 hours later when there was no active region present. This is done in order to reduce the changing projection effects as the active region is tracked across the disk. Except for the splitting into frequency intervals, no filtering was done. The lower-right panel shows the unsigned magnetic field averaged over the same patches to indicate the timeline of emergence. The locations of the selected patches are shown if Figure~\ref{Fig:Observation:Magnetogram}.}
	\label{Fig:Observation:TimeSeries}
\end{figure}

Encouraged by the simulation results, we have looked for evidence for this effect in observations.
Figure~\ref{Fig:Observation:TimeSeries} shows the result from our analysis of SOHO/MDI Dopplergrams of the emergence of active region NOAA~10488.
The observational data are split into three 1~mHz-wide frequency intervals, since different frequencies show different behaviors. 
In most of the frequency intervals, the power inside the patch where the region emerges and outside stayed very much the same until about 300~minutes or so into the time series starting at 06:00:00 UT, 26 October 2003.
This is slightly after the time when significant magnetic flux was
first visually seen on the surface, although the magnetic field actually starts to appear much earlier at around 200~minutes. 
The frequency interval between 3 and 4~mHz however shows a significant difference in acoustic power before the emergence.
Here, the two curves for the quiet Sun and the patch where the active region emerges diverge before significant magnetic flux appears on the surface.
The power inside the patch was lower than outside, which is, according to the numerical simulations, compatible with the presence of a subsurface region with reduced wave speed.
Why a significant signal is only seen for frequencies between 3 and 4~mHz is currently not understood, but it is consistent with the simulations in which this frequency band showed by far the strongest variations in acoustic power when subsurface regions were present. 

\begin{figure}
	\centering
	\vspace*{0.3cm}
	\includegraphics[width=4cm,angle=90]{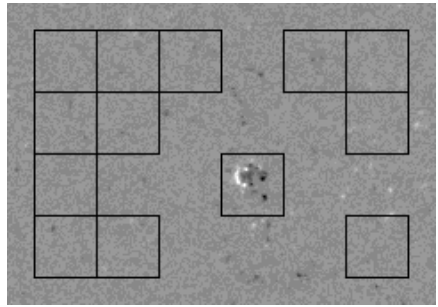}
	\vspace*{0.3cm}
	\caption{Magnetogram during the emergence of NOAA active
          region 10488 at 350~minutes into the time series. The square
          outlines indicate the patches used in
          Figure~\ref{Fig:Observation:TimeSeries}. The central region
          is the patch were the active region is emerging, and the quiet-Sun patches around were used for comparison.}
	\label{Fig:Observation:Magnetogram}
\end{figure}                 

We have looked at a few other emergence events for which Dopplergrams with one-minute cadence are available from SOHO/MDI.
The results for these have been inconclusive, though, in part due to data quality. 
The above analysis uses very short time series of 128 minutes and small patches, and therefore the noise level is relatively large to begin with.
The signal is very sensitive to missing or damaged images (even interpolated). 
We found that even short gaps may cause significant variations in the acoustic power signals, confusing the results.

\section{Conclusions}

Numerical simulations of solar acoustic wave propagation show that under certain conditions subsurface structures that modify the propagation speed of the waves result in, depending on the sign of the perturbation, a reduction, for negative perturbations, or enhancement, for positive perturbations, of the acoustic power observed at the photosphere above them.
The power variation is strongest for frequencies between 2 and 4 mHz.
There is observational evidence that, at least for the emergence event shown above, a reduction in acoustic power for frequencies between 3 and 4 mHz can be seen before significant magnetic field is visible at the surface.
This is compatible with the presence of a subsurface structure with reduced wave speed.
Further investigation of the acoustic power signals associated with emerging active regions may provide important constraints on their properties, such as the size, depth structure, and emergence speed.

%

%

%
\bibliographystyle{spr-mp-sola}
\bibliography{hartlep}  

\end{article} 
\end{document}